\documentstyle[12pt]{article}

\begin{document}
\begin{titlepage}
\title{Weibel instability on weakly relativistic produced plasma by circular polarization microwave electric field
\footnote{PACS52.40.Db,52.35.Hr}}
\author{M.~Ghorbanalilu\footnote {E-mail:m-alilo@cc.sbu.ac.ir}\\{\small \it
Islamic Azad Tabriz University, Tabriz, Iran}\\B. Shokri\\
{\small \it Physics Department of Shahid Beheshti University,
Evin, Tehran,Iran}}
\end{titlepage}
\date{}
\maketitle
\begin{abstract}
Analyzing the production of weakly relativistic plasma produced by
microwave fields with circular polarization the electron
distribution function is obtained to be non-equilibrium and
anisotropic. Furthermore, it is shown that produced plasma is
accelerated on the direction of propagation of microwave electric
fields. The electron velocity on this direction strongly depends
on electron origination phase, electric field phase, and
amplitude of microwave electric field. Making use of the
dielectric tensor obtained for this plasma, it is shown that the
wiebel instability develops due to the anisotrpic property of
distribution function. The dispersion equation is obtained for
this instability and the growth rate of it, is calculated.
\end{abstract}
\newpage \vskip 1cm
{\bf\large I. INTRODUCTION} \vskip 0.5cm

 A microwave produced gas discharge is a rather complicated phenomenon
 exhibiting a variety of features. Numerous
theoretical and experimental studies have been devoted to this
phenomenon. The interaction between intense (MW) fields and
neutral gas open new possibilities for studying the fundamental
property of this produced plasma. In a plasma produced in the
interaction of the MW field with a gas, the electron distribution
is non-equilibrium and may give rise to various plasma
instability.$^1$ The main features of this interaction is the gas
ionization mechanism. Furthermore, the key role in the plasma
processes is played by the kinetic effects associated with the
specific features of the electron distribution. Due to the short
interaction times, these features are governed completely by the
pulse parameter of microwave field. In addition the interaction
between intense MW fields with an inhomogeneous plasma results in
variety of phenomena such as frequency up shift during the
propagation of a pulse microwave field through a plasma.$^2$ In
such a strong wave field the electron oscillatory energy~
$\epsilon_e$ is much higher than the ionization energy~$I_{ioniz}$
of gas atoms
    \begin{equation}\label{1}
    \epsilon_e=\frac{e^2E_0^2}{2m\omega_0^2}\gg I_{ioniz},
    \end{equation}
where ~$\omega_0 $~is the radiation frequency, $m$ is  electron
mass,and~ $E_0$ is the electric field strength. When field
amplitude is comparable to the atomic field
strength~$E_a\approx5.1\times 10^9~ Vcm^{-1}$ the tunneling
ionization becomes an important mechanism for direct ionization
of the gas atoms. This effect was completely studied in the
different papers.$^{3,4,5}$ On the other hand relativistic
effects come in to play when kinetic energy of electron
oscillation in an electromagnetic field is comparable with the
electron rest mass energy. Moreover, the microwave fields
generated by the present day pulsed duration sources are weaker
than the atomic field and are capable of manifesting weakly
relativistic behavior of electron produced. On the oder side, the
aforementioned effect can be easily manifested by intense laser
pulse.$^6$

In the previous paper we restricted our study to the
non-relativistic regime but here we consider weakly relativistic
effects that occur during gas ionization by a strong microwave
field.$^7$ In the present paper we consider the interaction of
circularly polarized microwave pulse fields with frequency
about~$\omega_0\simeq 2\times10^{10} s^{-1} -2\times10^{11}
s^{-1}$ with a neutral gas taking into weakly relativistic
effects. This pulsed radiation source is capable of generating the
radiation with an intensity of about $10^8Wcm^{-2}$, whose
electric field being~$ E_0 \leq 10^6 Vcm^{-1}$~ is much weaker
than the atomic field $E_a$. In weakly relativistic case $v_E \ll
c $, for example in the weak ionized gas at low plasma density,
the electron average drift velocity $v_{z av}$ arises in the wave
propagation direction. At $\omega_0\approx 2 \times 10^{10}s^{-1}
$ and $E_0\approx 10^4 V cm^{-1}$ we get $v_E\approx 0.9\times
10^9 cm s^{-1}$ and $v_{z av}= 2.6\times 10^7 cm s^{-1}$. If
plasma density could get its critical value the drift current in
plasma ought be $j_z\approx 0.5 Acm^{-2}$. $^8$

In this case, also we study the electron distribution function
(EDF) and the stability of the discharge plasma in the
aforementioned frequency range at weakly relativistic electron
oscillation energy~$\epsilon_e$~.

This work is organized in four section. In sec. II. we will
obtain produced EDF generated by the interaction of the
circularly-polarized MW field with a neutral gas. In sec.III. we
will obtain the dielectric tensor element for this produced
plasma. In sec.IV. we will study the stability of produced plasma
and find the wiebel instability growth rate of wiebel instability
caused by the anisotropic property of the electron distribution
function. Finally, a summary and conclusion is presented.

    \vskip 1cm {\bf\large II. ELECTRON DISTRIBUTION FUNCTION }\vskip 0.5cm

   Under condition~(1)~the thermal velocity of the electrons in a
discharge plasma can be neglected in comparison to the electron
oscillation velocity in the MW radiation field. Since the
collision frequency is much smaller than the MW field frequency,
we can ignore the collisional stochastization of the forced
electron oscillation as well. Furthermore, if plasma
density~$n_e(t)$  produced by the field during gas breakdown, is
less than  the critical density (that is~$\omega_0^2 >
\omega_{Le}^2=4\pi ne^2/m $) we can neglect the effect of the
polarization field. Moreover, the plasma density is assumed to be
less than the neutral gas density~$n_0$~so that the latter can be
considered constant. We also suppose that the field was
adiabatically  switched on in the infinite past. Furthermore, we
can assume that the MW radiation electric field amplitude ~${\bf\
E_0}$ is constant during a single field period. Therefore, the
kinetic equation for the plasma electrons produced in the gas
breakdown by a strong pulsed field can be written as follows
 \begin{equation}\label{2}
 \frac{\partial f_0}{\partial t}+{\bf\ v}.\frac{\partial
 f_0}{\partial {\bf\ r}}+e[{\bf\ E_0} + \frac{1}{c}({\bf\ v} \times
 {\bf\ B_0})].\frac{\partial f_0}{\partial {\bf\ p}}=
 n_0 \omega_{ioniz}  \delta( {\bf\ p}),
 \end{equation}
where~${\bf\ E_0}(\xi)$ and~$ {\bf\ B_0}(\xi)$ are electric and
magnetic fields of a wave propagating along the Z-axis,
respectively;~$ \xi=\omega_0 t-k_0 r=\omega_0(t-z/c)$, and
$\omega_{ioniz}$ is the ionization probability of the gas atoms.
Here $\delta({\bf\ p})$ is the delta function of the electron
momentum. In the weakly-relativistic limit, we can assuming that
the electric field depends only on time. In the case of MW
breakdown, electron-impact ionization is governed by the
ionization probability $\omega_{ioniz}$. For a MW discharge Eq.
(2) for $f_0({\bf\ p},t)$ is a homogeneous integro-differential
equation whose positive eigenvalue ~$ \gamma({\bf\ E_0})$
determines the avalanche ionization constant. However, we can
neglect the right-hand side of Eq. (2) in the first approximation
and calculate the EDF directly by solving the Vlasov equation
under the following condition,
\begin{equation}\label{3}
\omega_0\gg\gamma({\bf\ E_0}),\omega_{ioniz}
\end{equation}
where the avalanche ionization constant~$\gamma({\bf\ E_0})$~is to
be determined.$^8$ In this case, the condition (3) depends
strongly on the neutral gas density and is well satisfied at gas
pressures of~$ p_0\simeq 10-100$ Torr. In this approximation, to
calculate the electron energy distribution function, we assume
the field components to be circularly polarized
\begin{equation}\label{4}
  {\bf\ E_x}={\bf\ E_0} \sin \omega_0 t,
  ~~~~~~~~~~{\bf\ E_y}={\bf\ E_0 }\cos \omega_0 t,
~~~~~~~~~~~~{\bf\ E_z}=0,
\end{equation}
where, $\omega_0$ is the frequency of MW field. Moreover, electric
field amplitude ${\bf\ E_0}$ describes the slowly varying (over
the field period) microwave pulsed envelope. consider the
condition (3) and solving Eq. (2) by characteristic method the
equation of motion for electrons can be obtained as follows
\begin{equation}\label{5}
\left\{
\begin{array}{c}
\displaystyle m\frac{dv_x}{dt}=eE_x,
\\[5mm]
\displaystyle m\frac{dv_y}{dt}=eE_y,
\\[5mm]
\displaystyle m\frac{dv_z}{dt}=\displaystyle
\frac{e}{c}(v_xE_x+v_yE_y),
\end{array}
\right.
\end{equation}
from this we find solution of the vlasov kinetic
equation-characteristics
\begin{equation}\label{6}
\left\{
\begin{array}{c}
v_x=-v _E(\cos\varphi-\cos\varphi_0),
\\[5mm]
v_y=v_E(\sin\varphi-\sin\varphi_0),
\\[5mm]
v_z=\frac{v_E^2}{c}(1-\cos(\varphi-\varphi_0)),
\end{array}
\right.
\end{equation}
where, $\varphi=\omega_0t$ is the MW electric  field phase and
$\varphi_0=\omega_0t_0$ is the MW electric field phase when the
electron originates with zero momentum at time $t_0$,
$v_E=eE_0/m\omega_0 $ is the electron oscillatory velocity in an
alternating electric field. For the solution of the form
$f_0(v,t)=n_e(t)\widetilde{f_0}(v)$ we obtain$^7$\\

\begin{equation}\label{7}
\widetilde{f_0}(v_x,v_y,v_z)=\delta(v_x+v_E(\cos\varphi-\cos\varphi_0))
\delta(v_y-v_E(\sin\varphi-\sin\varphi_0))
\end{equation}
$$\times  \delta \left(v_z-\frac{v_E^2}{c}[1-\cos(\varphi-\varphi_0)]\right).$$
 The function ~$\tilde{f_0}(v_x,v_y,v_z)$ satisfies the normalization
condition $\int d{\bf\ v}~\tilde{f_0}({\bf\ v})=1$. One can show
the product of the two last $\delta$-functions in the distribution
function (7) is independent of $\varphi_0$. For this reason, by
introducing the following notation
$$ V_x=-v_y+v_E \sin\omega_0 t~,~~~~~~~~~~~~V_y=v_x+v_E\cos\omega_0
t~,$$ we obtain
$$\delta(v_x+v_E(\cos\omega_0 t-\cos\varphi_0))
\delta(v_y-v_E(\sin\omega_0
t-\sin\varphi_0))=\displaystyle\frac{1}{2\pi v_E}~
\delta(V_\bot-v_E)~,$$ where
$$ V_\bot^2= V_x^2+ V_y^2 =v_\bot^2+v_E^2+2 v_E(v_x\cos\omega_0
t-v_y\sin\omega_0 t ),~~~~~~~v_\bot^2=v_x^2+v_y^2~.$$ Therefore,
Eq. (7) is reduced to
\begin{equation}\label{8}
\tilde{f_0}(V_\bot,v_z)=\frac{1}{2 \pi v_E} \delta(V_\bot-v_E)
\delta\left(v_z-\frac{v_E^2}{c}[1-\cos(\varphi-\varphi_0)]\right).
\end{equation}
We can see that, the electron distribution function depends on
phase field $\varphi$ and electron origination phase $\varphi_0$.
Therefore, we must average of electron distribution function over
the phase field period and origination phase. The projection of
the phase portrait of the electrons onto the $(v_x,v_y)$ plane is
as follows. The electron trajectories uniformly cover a circle of
radius $v_E$, whose center precesses about the origin of the
coordinates and describes a circumference of the same radius at a
rate equal to the microwave frequency. Consequently, the
averaging procedure can be performed separately for the
longitudinal and transverse velocity components. Averaging over
the transverse component we obtain
\begin{equation}\label{9}
\widetilde{f_0}(v_\bot,v_z(\varphi-\varphi_0))=
\frac{\delta\left(v_z-\frac{v_E^2}{c}[1-\cos(\varphi-\varphi_0)]\right)}{2\pi^2
v_\bot\sqrt{4v_E^2-v_\bot^2}}
\end{equation}

\vskip 1cm {\bf\large III. DIELECTRIC TENSOR}\vskip 0.5cm

In the previous section we found the distribution function (9) for
electrons in a discharge plasma is highly anisotropic with
respect to the direction of the MW radiation field, which, first
of all, should result in the onset of the well-known wiebel
instability.$^7$ In order to convince ourselves that this
conclusion is valid and to find the instability growth rate, we
turn to the adiabatic approximation, assuming that the
instability grows faster than the plasma density.$^{5,9}$ In this
approximation, we can use the following dispersion relation for
small perturbation
\begin{equation}\label{10}
\left. \mid k^2\delta_{ij}-k_i
k_j-\frac{\omega^2}{c^2}\varepsilon_{ij}(\omega,\mathbf{k})\right.\mid=0.
\end{equation}
Here~$\varepsilon_{ij}(\omega,\mathbf{k})$,~the dielectric
permitivity tensor of the plasma, is obtained by linearizing
Vlasov equation for the electrons
\begin{equation}\label{11}
\frac{\partial f_e}{\partial t}+ {\bf\ v}. \frac{\partial
f_e}{\partial {\bf\ r}}+ e \left.({\bf\ E_0}+ \frac{1}{c} ({\bf\
v} \times {\bf\ B_0})\right.). \frac{\partial f_e }{\partial {\bf\
p}}=0,
\end{equation}
and considering self-consistency effect and Maxwell equations and
cold ions approximations as follows
\begin{equation}\label{12}
\varepsilon_{ij}(\omega,{\bf\
k})=1+\delta\varepsilon_{ij}^I+\delta\varepsilon_{ij}^e=1-\frac{\omega_{Pi}^2}{\omega^2}
\delta_{ij}+\frac{\omega_{Pe}^2}{\omega^2}\int d{\bf\ v}
\left[v_i\frac{\partial\tilde{f_0}}{\partial v_j}+ v_i v_j
 \frac{{\bf\ k}.\displaystyle \frac{\partial \tilde{f_0}}{\partial
{\bf\ v}}}{\omega-{\bf\ k}.{\bf\ v}}\right].
\end{equation}
Here, the second term, i.e.,~$ \delta\varepsilon_{ij}^I$~shows the
ion contribution of the plasma dielectric permitivity tensor and
the third term, i.e.,~$ \delta\varepsilon_{ij}^e$~is related to
the electron contribution of  the plasma dielectric permitivity
tensor. The latter kinetic equation for the perturbed EDF is valid
in the adiabatic approximation when short-wavelength limit is
justified, i.e., $ \omega_{Pi} \ll kv \leq \omega_{Pe}$.
 To solve Eq. (12) by using EDF (9) we introduce the following system of
coordinate in which~$\mathbf{k}$~lies only in the XZ plan
$$\mathbf{k}=(k_\bot,0,k_z)~~,~~~~~~~~~~~~k_\bot=\sqrt{k_x^2+k_y^2}.$$
Under this condition the non zero contributions of the electrons
in the dielectric tensor (12) are
$$\delta\epsilon_{11}=-\frac{\omega_{Pe}^2}{\omega^2}\left[1-\frac{k_z^2}{2k_\bot^2}
-\frac{(1-\frac{2k_z^2}{k_\bot^2})s}{2\sqrt{1-\frac{4k_\bot^2v_E^2}{\beta^2\omega^2}}}
+\frac{(1-\frac{k_z^2}{k_\bot^2})s}{2(1-\frac{4k_\bot^2v_E^2}
{\beta^2\omega^2})^\frac{3}{2}}\right] $$
$$\delta\epsilon_{33}=-\frac{\omega_{Pe}^2}{2\omega^2}\left[1-\frac{2i
k_\bot v_E^3(1-\cos\phi)^2}{\omega c^2\beta}(1+\frac{ik_\bot v_E
s}{2\omega \beta \sqrt{1-\frac{4k_\bot^2 v_E^2}{\omega^2
\beta^2}}}) - \frac{k_\bot^2 v_E^4(1-\cos\phi)^2}{\omega^2 c^2
\beta^2(1-\frac{4k_\bot^2 v_E^2}{\omega^2\beta^2})^\frac{3}{2}}
\right.$$
\begin{equation}\label{13}
\left.- \frac{2k_z v_E^2(1-\cos\phi) s}{  \omega c \beta
\sqrt{1-\frac{4k_\bot^2 v_E^2}{\omega^2 \beta^2}}}+\frac{k_z^2
v_E^4(1-\cos\phi)^2 s}{\omega^2 c^2 \beta^2(1-\frac{4k_\bot^2
v_E^2}{\omega^2\beta^2})^\frac{3}{2}}\right]
\end{equation}
$$ \delta\epsilon_{13}=\frac{\omega_{Pe}^2}{2 \omega^2} \left[\frac{ k_\bot v_E^2 (1-\cos\phi) s }
{\omega c
\beta}(1+\frac{k_z^2}{k_\bot^2})\left(\frac{1}{\sqrt{(1-\frac{4k_\bot^2
v_E^2}{\omega^2\beta^2})}}-\frac{1}{(1-\frac{4k_\bot^2
v_E^2}{\omega^2\beta^2})^\frac{3}{2}}\right)\right.$$
$$\left.+\frac{k_z}{k_\bot}(1-\frac{s}{\sqrt{1-\frac{4k_\bot^2
v_E^2}{\omega^2 \beta^2}}})\right]$$
$$\delta\epsilon_{12}=\delta\epsilon_{21}=0,~~~~~~~\delta\epsilon_{22}=-\frac{\omega_{Pe}^2}{\omega^2}~~~~~~~~~
\delta\epsilon_{31}=-\frac{i\omega_{Pe}^2
v_E}{2\omega^2c}(1-\cos\phi)+\delta\epsilon_{13},$$
 where
\begin{equation}\label{14}
s=\left\{
\begin{array}{c}
\left.
\begin{array}{c}
 2~~~~~~for~~~~~~~~~~~~~~~~~~~ Re\omega\neq0\\
 1~~~~~~for~~~~~~~~~~~~~~~~~~~Re\omega=0
\end{array}\right\}when~~~~~~\frac{4k_\bot^2 v_E^2}{\beta^2\omega^2}<1
\\[8mm]
\left.
\begin{array}{c}
0~~~~~~for~~~~~~~~~~~~~~~~~~~ Im\omega=0\\
2~~~~~~for~~~~~~~~~~~~~~~~~~~ Im\omega\neq0\\
1~~~~~~for~~~~~~~~~~~~~~~~~~~Re\omega=0
\end{array}\right\}when~~~~~~\frac{4k_\bot^2v_E^2}{\beta^2\omega^2}>1
\end{array}
\right.
\end{equation}
and $\beta=(1-\frac{k_z v_E^2}{\omega c}-\frac{k_z v_E^2}{\omega
c} \cos\phi)$ , $\phi=\varphi-\varphi_0$ also, $i=\sqrt{-1}$.

\vskip 1cm {\bf\large IV. STABILITY OF THE PRODUCED PLASMA}\vskip
0.5cm

As known the process of the gas breakdown by the  MW fields is
unstable with respect to the excitation of longitudinal electric
fields and transverse magnetic fields. The former is caused  due
to the positive derivative of the EDF and the latter, treated
below, is related to the anisotropy of the EDF.$^{9}$ In this step
we study the weibel instability,  by analyzing  electron
perturbations propagate across the microwave field. Therefore, by
averaging  dielectric tensor (13) over the $\phi$ between
$[0,2\pi]$ and substituting into the dispersion equation (10)
 by assuming $2kv_E^2/\omega c <1$ yields the dispersion relation for electron perturbations
propagating across the MW radiation field ($k_\bot=0$)
\begin{equation}\label{15}
\omega^3+\frac{\omega_{Pe}^2k^2v_E^2\omega}{(k^2c^2+\omega_{Pe}^2)}
+\frac{2\omega_{Pe}^2k^3v_E^4}{c(k^2c^2+\omega_{Pe}^2)}=0.
\end{equation}
 With assumption $k^2v_E^2\ll\omega_{Pe}^2\ll k^2c^2$ and Solving Eq. (15) we obtain the following expression, which
characterize the growth rate of the weibel instability as follow
\begin{equation}\label{16}
\omega = \frac{k^2v_E^2}{\sqrt{k^2c^2+\omega_{Pe}^2}}+
i\frac{kv_E\omega_{Pe}}{\sqrt{k^2c^2+\omega_{Pe}^2}}\left(1+\frac{3k^2v_E^2}{2\omega_{Pe}^2}\right)
\end{equation}

\vskip 1cm {\bf V. CONCLUSION }\vskip 0.5cm

Using a simplest model, we are calculated the weakly-relativistic
EDF for plasma produced by the interaction of an intense
microwave pulse field with a neutral gas. The resulting EDF, which
can be drive analytically for circularly polarized MW field, is
highly anisotropic, which indicates that microwave driven plasmas
are subject to weibel instability. In the non-relativistic
produced plasma$^7$ the electrons only oscillate on the MW field
direction but on the weakly-relativistic produced plasma the
electrons have velocity perpendicular the radiation field. In
this case the electrons originating with zero energy are
entrained by MW field in a certain phase and are accelerated to
weakly-relativistic velocities, so that the accelerated electrons
move essentially in phase with the MW field. Then, this process
might be repeated for the electrons originating over the next time
interval, and so on. Analyzing the dispersion equation and
obtaining the growth rate of the instability produced we found
that the growth rate modified by an exceed positive term with
respect to non-relativistic case. Also the instability frequency
has a very small real part that can be represent the entrain of
electrons in MW fields.

\end{document}